\documentclass[twoside]{article}
\usepackage[sc]{mathpazo}
\usepackage[T1]{fontenc} 
\usepackage{microtype}
\usepackage{multicol}
\usepackage{graphicx}
\usepackage{listings}
\usepackage{amssymb,amsfonts,amsmath}
\usepackage[hmarginratio=1:1,top=32mm,columnsep=20pt]{geometry} 
\usepackage{hyperref}
\usepackage{lettrine} 
\usepackage{textcomp}
\usepackage{abstract}
\usepackage{color,listings}
\definecolor{lightgray}{rgb}{.7,.7,.7}
\definecolor{white}{rgb}{1,1,1}
\newsavebox\lstbox

\usepackage{fancyhdr}
\title{\vspace{-12mm}\fontsize{24pt}{25pt}\selectfont\textbf{Population Dynamics of Self-Replicating Cell-like Structures Emerging from Chaos}}
\author{
\large
\textsc{Thomas Schmickl\footnote{Corresponding author: \href{mailto:thomas.schmickl@uni-graz.at}{thomas.schmickl@uni-graz.at}}} \ and \textsc{Martin Stefanec}\\[2mm] \\ 
\small Department of Zoology, Karl-Franzens University Graz, Austria. \\ \small Universit\"atsplatz 2, A-8010 Graz, Austria}
\date{\small submitted $14^{th}$ December 2015}

\begin{document}

\maketitle 

\thispagestyle{fancy} 

\begin{abstract}

We present here a system of self-propelled particles that follow a very simple motion law in continuous space in a deterministic and asynchronous way. This system of particles is capable of producing, depending on the particle density in the habitat, several spatio-temporal patterns emerging from an initial randomized spatial configuration. We found that those structures show specific population dynamics which arise from death (decay) and growth (self-replication) of those structures, thus we call the system Primordial Particle System (PPS), as the model can be interpreted as a simplistic model of emergence of self-replicating chemical structures from initially chaotic mixed components in the ``primordial soup'' at the beginning of life. We describe the observed dynamics, show the emerging spatio-temporal structures and present a macroscopic top-down model as well as a probabilistic microscopic bottom-up model of the system. \vspace{2mm}

\setlength{\parindent}{0pt}Keywords: Self-organization, Emergent pattern formation, Self-replication, Protocell model, Artificial Life, Emergence of Life, Morphogenesis  \vspace{2mm}

\end{abstract}

\begin{multicols}{2} 

\section{Introduction}

At the very beginning of life we assume a biological singularity: Based on simple local interaction rules initially randomly distributed particles (molecules) spontaneously self-organized into aggregates of higher order \cite{citation:1}. These compounds first managed to withstand the thermodynamic path towards entropy and sustained their orderedness over some time \cite{citation:2}. Very likely these forms represented an equilibrium state of decay (losing energy and/or losing elements to the environment) but also growth/repair (intake of energy and/or intake of elements from the environment)\cite{citation:1}\cite{citation:3}. Additionally, these patterns had to be self-reproducing, otherwise they would have vanished after some time. Very likely, at least at some point, these creatures must have built a barrier between their inner reactive core and the outside environment by forming a sort of proto-cell \cite{citation:4}\cite{citation:5}. 

In this study we present a simple model that demonstrates that such self-reproducing protocells can emerge spontaneously from a population of homogenous, purely reactive, deterministically moving (self-propelled) particles \cite{citation:6}\cite{citation:7} following one single very simple rule. Following this approach we looked for a system of minimum complexity of particle motion that is capable of exhibiting spontaneous emergent morphogenesis by forming self-reproducing structures from initial random chaos. Our model is capable of exhibiting such phenomena on a regular basis, thus we call our system ``Primordial Particle System'' (PPS), as it mimics the initial emergent morphogenesis in the primordial soup. In the following we first develop the model, then we showcase selected results and finally we report systematic analysis of some core properties of the system.

\section{The mathematical model}

A PPS models a population of particles moving deterministically and asynchronously in a continuous toroidal wrapped space. Each particle is defined by its position $\mathbf{p}_t = (x_t ,y_t )$, by its heading $\phi _t$ and by its movement with constant velocity $v$, thus $\frac{ \Delta \mathbf{p}}{\Delta t}=
\begin{pmatrix}
cos\phi\\
sin\phi\\
\end{pmatrix}
\cdot v$, assuming an open system allowing steady energy influx  for motion. An explanatory implemented PPS motion law as pseudo-code is provided at Listing \ref{code:example}. Particles react to other particles within radius $r$ by changing their heading: Turn directionality depends on particle numbers left ($L_{t,r}$) and right ($R_{t,r}$) of them within distance $ d_t \leq r $, while turn angles depend on the total number of surrounding particles ($N_{t,r}=L_{t,r}+R_{t,r}$). This yields the final motion law for PPS:

\begin{equation}
\label{eq:2}
\frac{\Delta\phi}{\Delta t}=\alpha+\beta\cdot\mathit{N_{t,r}\cdot sign}(R_{t,r}-L_{t,r})
\end{equation} where $\alpha$ represents a fixed rotation and $\beta$  models a rotation proportional to local neighborhood size.   

Neighborhood configuration affects $\phi_t$ in each timestep, in turn changing a particle\textquotesingle s position, ultimately yielding new local configurations. This feedback loop governs the self-organization of the PPS system. \\
With $\alpha$ $\mathit{=180^{\circ}}$, isolated particles hold position within 2 timesteps. Only when other particles enter their neighborhood they start to move away: PPS with $\alpha$ $\mathit{  =180^{\circ}}$ ``mirror'' the behaviour of particles in PPS with $\alpha$ $\mathit{=0^{\circ}}$ on the microscopic level.\\
Spatial implementation of the model: In our PPS each particle holds position $\mathbf{p}_t = (x_t, y_t)$ and heading $\phi_t$ at every time step t. The change of this heading ($\Delta \phi$) is modeled in our model equation \ref{eq:2}. Every positive heading change is considered to be a turn to the right (first person view), respectively clockwise (third person view), every negative $\Delta \phi$ is a left turn (first person view), respectively counter clockwise (third person view).  The positional update is modeled according to $\mathbf{p}_{t+1} = \mathbf{p}_t + ((cos \phi_t),  (sin \phi_t)) \cdot v$, where v is the constant velocity of each particle. The following pseudo-code describes the algorithmic implementation, where 0 degrees are assumed to be top-side:
\begin{lstlisting}[frame=single,language=Python,caption=Explanatory implementation of a PPS as pseudo-code \label{code:example}]
loop foreach timestep {
  loop foreach particle {
  L = (count left neighbors in 
  radius r)
  R = (count right neighbors in 
  radius r)
  N = L + R
  delta_phi = 
  alpha + beta * N * sign(R - L)
  turn-right delta_phi
  move-forward v
  }
}\end{lstlisting} We discovered an exceptionally interesting system with the parameter set $PPS=\langle r=5,\alpha =180^{\circ}, \beta=17^{\circ}, v=0.67 \rangle $, used as default PPS parameters here. For visualization we color-code particles by their local neighborhood size: $color_t = if \ 16<N_{t,r=5}\leq35 \ blue \ else \ if \ N_{t,r=5}>35 \ yellow \ else \ if \ N_{t,r=1.3}>15 \ magenta \ else \ if \\ 13 \leq N_{t,r=5}\leq15 \ brown \ else \ green$. 
These colors indicate an exhaustive classification of each particles rotational power. We used a density-dependent population model to interpret observed population dynamics of emerging structures \cite{citation:8}:
 $\Delta X/\Delta t=a\cdot(1-X/K)\cdot X$, where the variable $X$ models the population size, the constant $a$ resembles the maximum reproduction rate per step and the constant $K$ is the carrying capacity of the habitat (Fig. 3a). Best fit was found by applying the minimum residuals method using least squares.




\section{The study}
After we discovered the PPS in general and the especially interesting parameter set of $ PPS=\langle r=5,\alpha=180^{\circ},\beta=17^{\circ},v=0.67\rangle $ we investigated a set of research questions to understand the properties of the emerging phenomena. Our research questions were: \\

\setlength{\parindent}{0pt} Q1: Do ordered structures emerge from random initial particle configurations? \\
Q2: Do such populations of structures grow over time? \\
Q3: Is the system converging to a certain ratio between different structure types? \\
Q4: Do the observed populations of structures follow dynamics known from nature? \\

To investigate those questions we started such PPS with a particle density of 0.08 particles per space unit (this means 8000 particles in a space of 250 units by 250 units) distributed randomly (uniform random distribution) and with randomized initial heading. The PPS does not contain any stochastic component, thus it is a fully deterministic motion law. All particles move asynchronously in randomized order. We performed 9 such simulation runs and recorded all color transitions of particles which express significant changes in local density. We recorded the spatial evolution of the system (picture sequences) and used a previously determined average number of blue and yellow particles in ``cell structures'' (48 such particles per cell) and a previously determined average number of pink particles in ``spore structures'' (18 particles per spore) to record population dynamics of those structures. To compare populations of different type and times we used Mann-Whitney-U-tests after applying Bonferroni correction as we made 9 ad-hoc tests on our collected data to prevent false-positives (type-1 errors) due to multiple testing.
\section{Results}
Figure 1 shows the resulting distributions of particles after 100,000 timesteps in 9 runs each started from a random initial distribution. It is clearly visible that in all runs both typical structures (very dense magenta spores and extended cell-like structures in blue/yellow) emerged and stayed alive for this extended time span. In all runs those structures started by initially building a single spore somewhere at a random place in the environment. Afterwards this structure grew to a cell which then replicated into more cells or spores. After some runtime a population of such structures inhabits the whole habitat. These structures are areas of denser particle populations thus the density of particles outside of these structures goes down in consequence, as the system is a closed system material-wise. However, it is an open system energy-wise as particles are self-propelled. The lowered density of the free particles allows them to arrange in a rather regular hexagonal-matrix-like configuration. \\
Figure 2 shows that the estimated populations of cells and spores show typical logistic growth dynamics. A comparison of populations of cells and spores in the initial phase shows that the system produces significantly more cells than spores, a fact that we observe also for later points in time (Mann-Whitney U-test, $N_1=N_2=9$ runs, $p<0.001$ for all those comparisons). Comparing the populations of those structures in an initial phase (step 2000) to later periods (step 5000 and step 100,000) shows that there is an initial growth phase (Mann-Whitney U-test, $N_1=N_2=9$ runs, $p<0.001$ comparing consecutive periods). Later periods (in longer runs) show no significant differences in population size anymore (data not shown). Figure 3a shows that the observed populations of spores and cells can be fitted to the classical top-down model of sigmoid (density-dependent) growth of natural populations (see for example \cite{citation:9}) yielding a growth rate of $a=7.1 \cdot 10^{-4}$ per step for cells and $a=4 \cdot 10^{-4}$ per step for spores and carrying capacities of $K=18.21$ spores and $K=50.78$ cells. Figure 3b shows a microscopic observation of state transitions of particles, as the different colors indicate different states of local densities. In the growth phase of the population the majority of state changes happen from green (very low density) to brown (slightly higher density) to blue (medium density). After the growth phase of the population there is an almost balanced flow through brown $\rightarrow$ blue $\rightarrow$ magenta $\rightarrow$ brown and green $\rightarrow$ brown $\rightarrow$ blue $\rightarrow$ green, whereby blue indicates high particle density and magenta indicates very high particle density.

\section{Discussion and Conclusion}

Our general and overall results are the novel discovery that this simple motion law, which we call PPS, is capable to self-generate ordered structures (cells, spores) from initial randomized spatial configuration of particles. This is an emergent phenomenon of the system. Figure 1 shows that research question Q1 was affirmed in 9 out of 9 tested runs. We further demonstrate that the arising populations differ between spores and cells (Q2) and show significant growth (Q2) over time which saturates after free particles were consumed by structures so that a density-dependent equilibrium emerges (Fig. 2). After the growth phase we observe a clear ratio of spore populations to cell populations (Q3, Fig. 3). By fitting the observed data to a classical population model from biology we show that the structures follow known population dynamics of natural organisms (Q4, Fig.3) suggesting the emergence of a sort of self-generating ecosystem. This is further supported by identifying clear ``material cycles'' that emerge in the system, as they are also known from ecology (Fig 3).

Based on those findings we conclude that we found a novel model capable of capturing a rich set of phenomena known from natural systems with a minimal set of computational complexity. We here report our first findings and assume that further analysis of the system will yield more fascinating insights and phenomena. 

A very prominent model in Artificial Life research and research of self-organization, emergence and complexity was the ``game of life'' (\cite{citation:10}), which consists of cells of binary state (dead or alive) in a discrete grid world following 3 simple rules of local interaction. These rules mimic simple ecological rules (overpopulation, Allee effect, partner-finding in reproduction) and create a rich set of patterns that are attributed with distinct names e.g., called ``pond'', ``glider'', ``comb''. It was found that the populations of cells in the GOL develops into a balanced population density, what is not surprising, as the microscopic rules in the GOL mimic the basic mechanisms of density-regulated growth (\cite{citation:8}). 

We took such ground-breaking and classical models (Game of Life, logistic growth and self-propelled particles), which were extensively studied in tens of thousands of analysis and articles, as a reference and created a system here that is even closer to the singularity of the origin of life and the ecological equilibria emerging afterwards (ecology): Random particles moving in a continuous space and interacting in an asynchronous way, forming patterns of life-like shape and complexity, self-reproducing, self-sustaining, self-repairing and developing into a well-balanced ecology. Also in our model, we look for a minimalistic set of rules achieving a maximum of diversity of patterns that emerge. We think that PPS is currently the most simple bottom-up model operating in continuous space and without required synchrony of agents (PPS work synchronously and asynchronously, data not shown) that can demonstrate morphogenesis from scratch. Thus we propose PPS as the most simple protocell model.
This article is a preliminary version of an article currently under preparation to be submitted to a high ranking scientific journal.

\section{Acknowledgments}
This article was written in cooperation of T.S. and M.S.: T.S. discovered the PPS and the interesting parameter setting and implemented the first programs of those PPS. M.S. conducted the systematic numerical analysis and statistical evaluation and produced all graphs of this article. The text of this article was equally written together by T.S. and M.S.

Early variants of such minimalistic algorithm to produce coordinated group behaviours of agents were investigated by the EU-ICT (FP7) project CoCoRo \#270382. Currently we further explore those systems with the funding of  the EU-H2020 FET-PROACTIVE project subCULTron \#640967 and the project REBODIMENT (Austrian Science Fund, FWF), P 23943-N13.

\end{multicols}

\begin{figure}[p]
    \centering
    \includegraphics[width=1\textwidth]{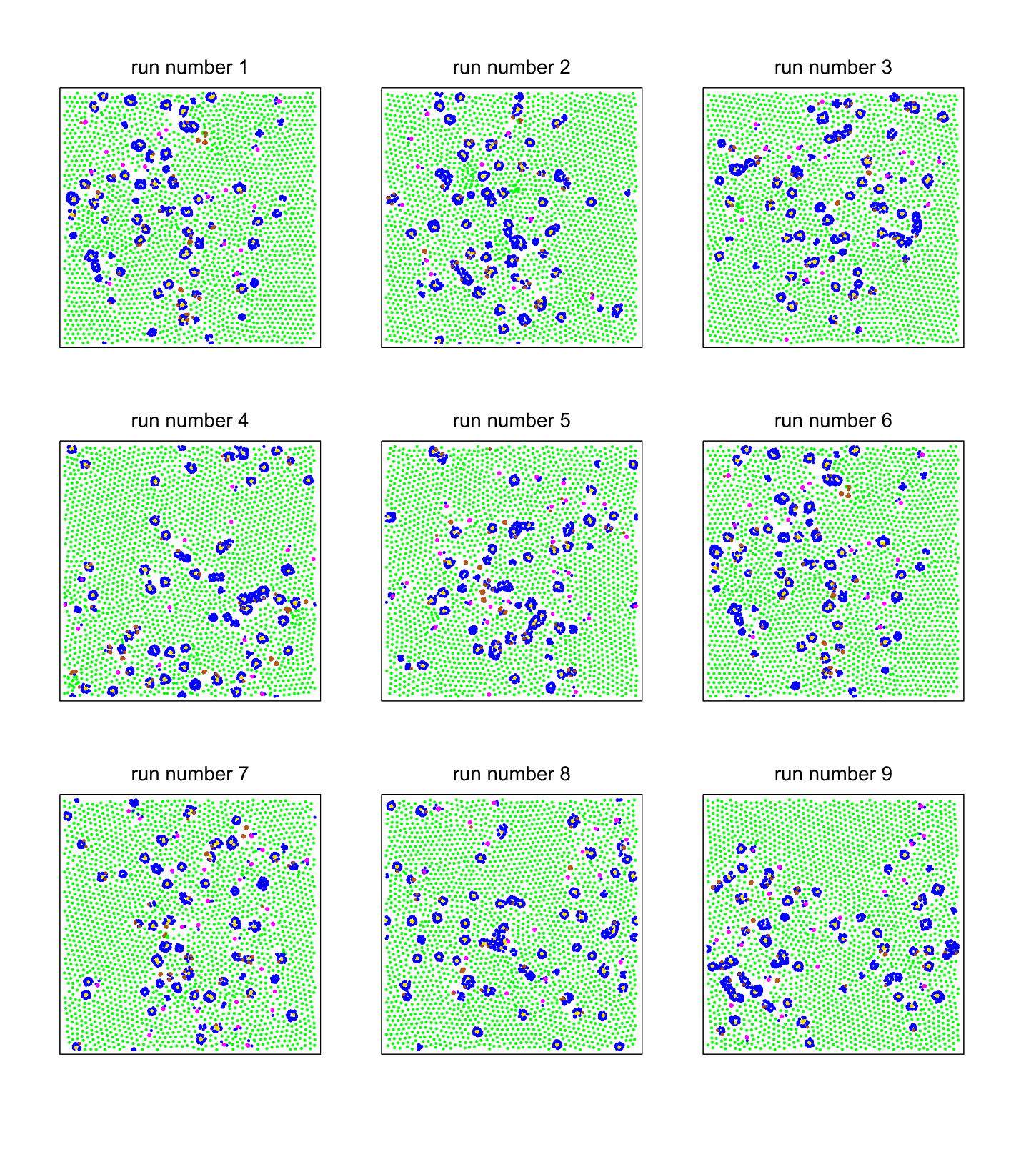}
    \caption{\textbf{Screenshots of 9 independent runs at the same timestep (100,000), at which color in visualization is determined by local neighborhood density.} Particles with more than 15 neighboring particles in a radius of 5 space units are colored in blue, while particles with more than 35 neighboring particles are colored in yellow. We call structures of blue and yellow particles ``cells''. Particles with more than 15 particles in a radius of 1.3 space units are colored in magenta. We call magenta structures ``spores''.}
    \label{fig:Figure1}
\end{figure}

\begin{figure}[p]
    \centering
    \includegraphics[width=1\textwidth]{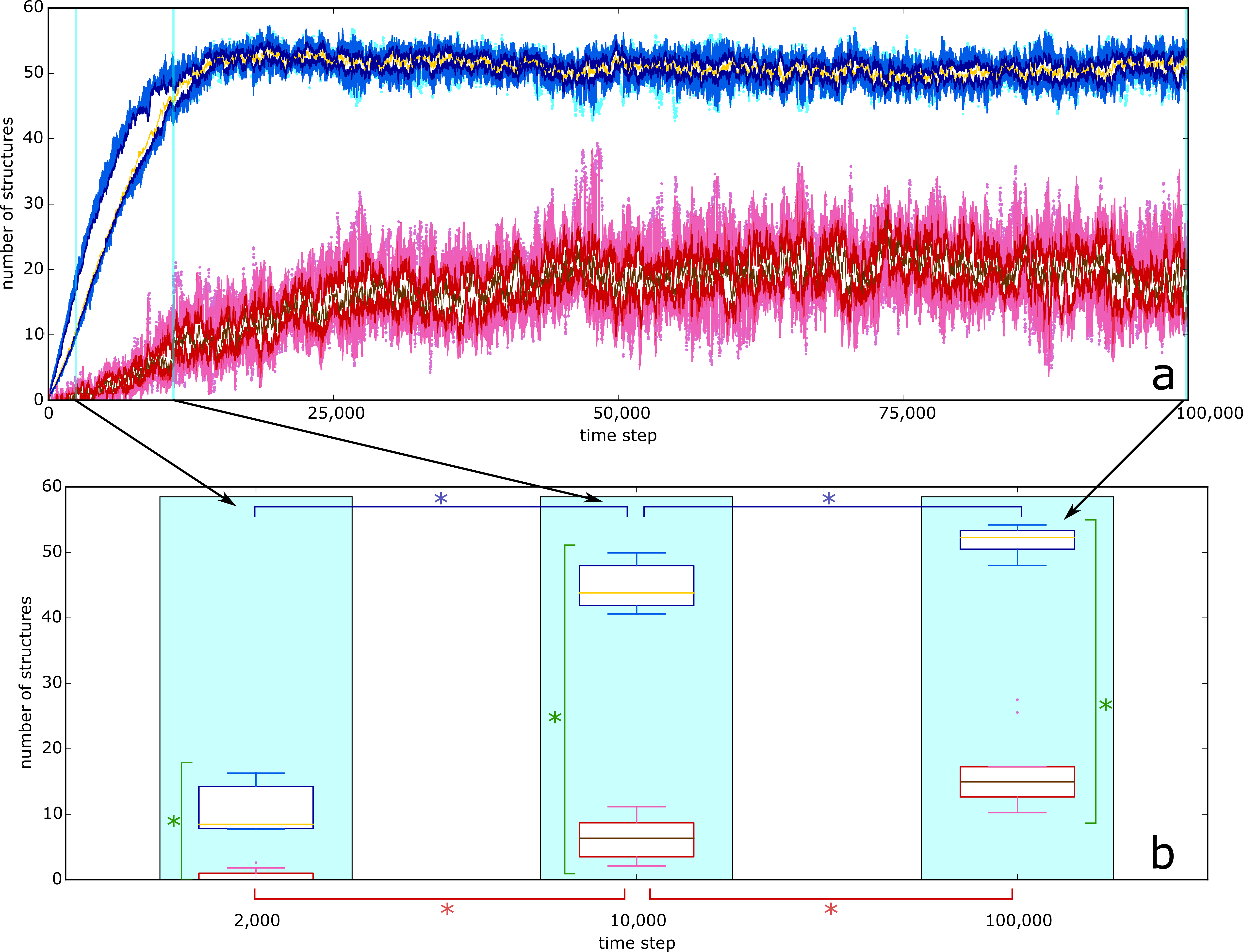}
    \caption{\textbf{Functional boxplot (a) number of cells (blue) and spores (red) at each timestep of 9 independent runs (seen in Figure 1), (b) example timesteps, statistically compared.} Figure 2a shows the number of cells at each timestep of 9 independent runs, depicted as outliers (cyan), 1.5 x IQR (dodger blue), first and third quartile (navy blue) and median (yellow). The number of spores at each timestep is characterized by outliers (very light pink), 1.5 x IQR (light pink), first and third quartile (red) and median (brown).  Figure 2b shows the populations queried at three exemplary chosen timesteps displayed as box plot. The number of cells (blue tone) and spores (red tone) between each other (green indicator) and the following timestep (red respectively blue indicator) were statistically compared using Mann-Whitney U-test ($N_1=N_2=9$ runs, $p<0.001$).}
    \label{fig:Figure2}
\end{figure}

\begin{figure}[p]
    \centering
    \includegraphics[width=1\textwidth]{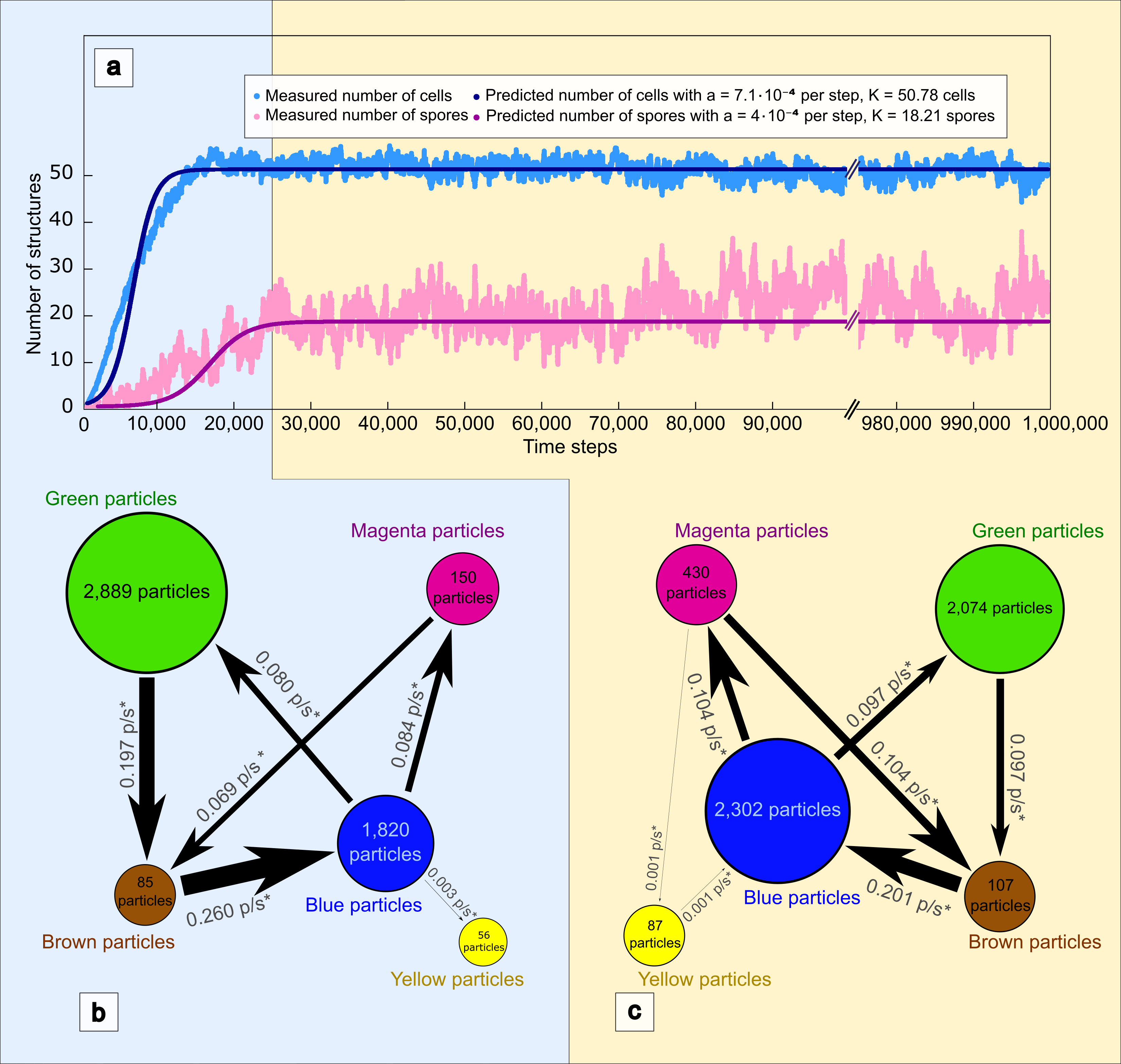}
    \caption{\textbf{Detailed analysis of one extended run (run number 5).} Figure 3a: Number of cells and spores at each timestep with fitted logistic growth model. Figure 3b and 3c: Analysis of microscopic movement of colored particles in step 0 to step 25,000 (3a) and step 25,000 to step 1,000,000 (3b). Figure 3c: Number of cells and spores at each timestep with fitted logistic growth model   $\Delta X/\Delta t=a\cdot(1-X/K)\cdot X$. These observed rates can be used for probabilistic Markov model of the system} 
    \label{fig:Figure3}
\end{figure}

\clearpage 

\end{document}